\documentclass[preprint,12pt]{elsarticle}

\usepackage{graphicx}
\usepackage{amssymb}
\usepackage{lineno}

\usepackage{txfonts}
\usepackage[english]{babel}
\usepackage{lipsum}  % Generates dummy text
\usepackage[hidelinks]{hyperref}  % Allows you to click on references in the PDF.
\usepackage{float}

\journal{ArXiv}

\begin{document}

\begin{frontmatter}


\title{Tracking the \textit{untracked}}



%----------------------------------------------------------------------------------------
%	TITLE
%----------------------------------------------------------------------------------------

\author{Aryabrata Basu\corref{cor1}}
\ead{aryabrata.basu@emory.edu}
\cortext[cor1]{Corresponding author}
\address{Emory University}
\address{Atlanta, Georgia, United States}

\date{\today} % Leave empty to omit a date

\begin{abstract}
\textit{The issue of seamless identification of users previously tracked using existing real-time optical position tracking system such as the OptiTrack system and maintaining continuous tracking state (history) of each of those users is a hard problem. In this article, we present a theoretical framework to integrate existing tracking systems with features such as user identification and history of up to `n' person activity. In our approach, we assume no direct communication with the tracking system, but access to all data it collects. Also, there are no guarantees that 1) the order of each tracked retro-reflective sphere reported is the same, and 2) that there will be any particular number of spheres in the room at any given time. We describe how the data is fused with existing tracking data to provide a seamless transition between other forms of position tracking}. 
\end{abstract}


\end{frontmatter}

%----------------------------------------------------------------------------------------
%	ARTICLE CONTENTS
%----------------------------------------------------------------------------------------

\section{Introduction}

\begin{figure}[t]
 \centering
 \includegraphics[width=\columnwidth]{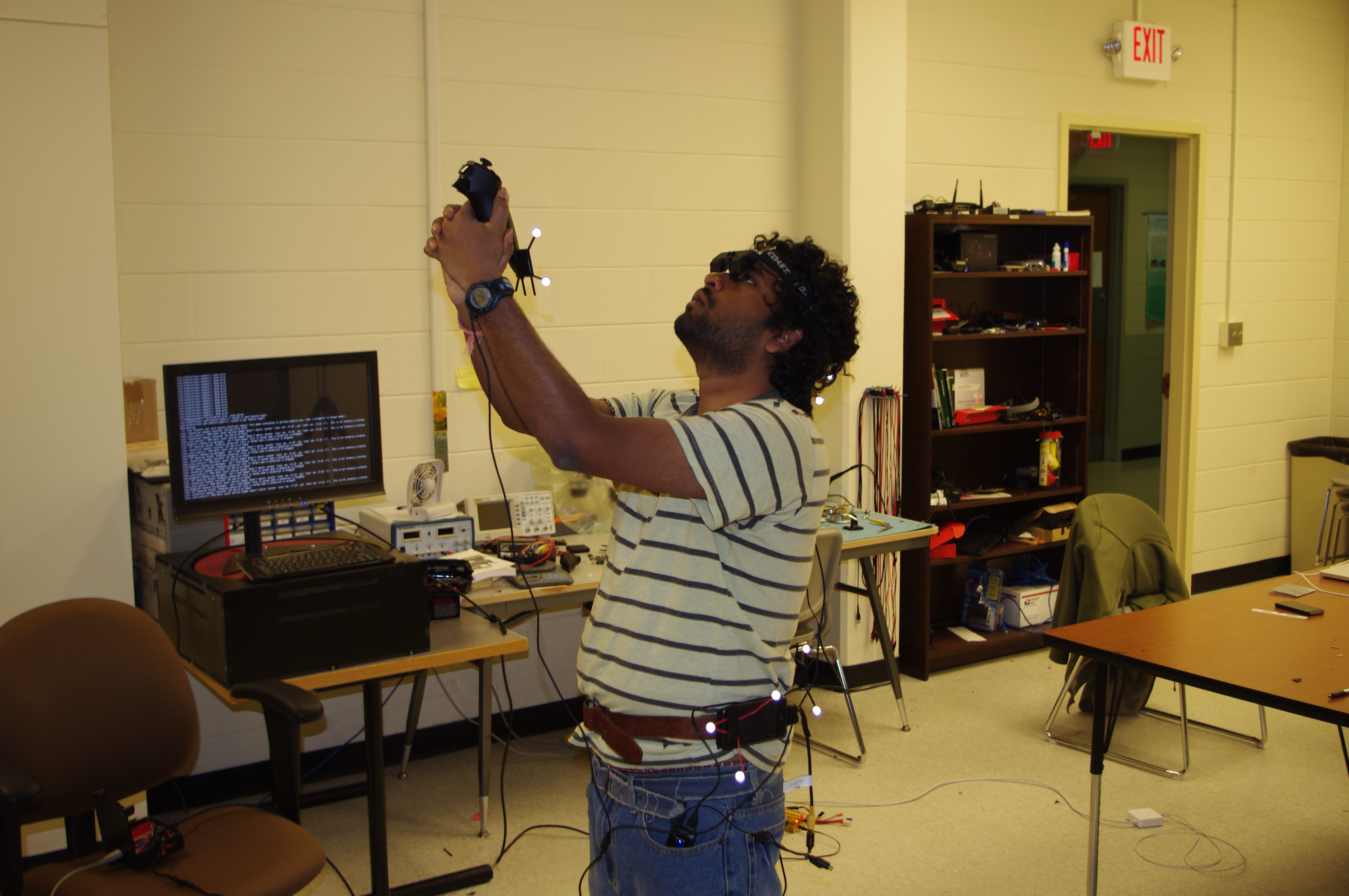}
 \caption{This is an example of a real-time position tracking using by a five camera OptiTrack system (Flex 3 cameras) with retro-reflective markers being tracked at 100 FPS. This picture is a courtesy of the Virtual Experiences Lab at University of Georgia.
}
 \label{fig:trackingvel}
\end{figure}

Virtual Reality (VR) as a medium allows its participants to experience an embodied perspective \cite{sherman2002understanding}. For example, in a flight simulator, the user embodies a virtual flight through direct control of a virtual cockpit. In order to elicit a perfectly immersed virtual cockpit, the \textit{VR system} needs to track the user's head gaze and synchronize the ego-centric perspective to match the user's head gaze. This is a form of sensory feedback by a VR system. Sensory feedback is essential to having a VR experience and a VR system provides direct sensory feedback to the user based on their physical position [\autoref{fig:trackingvel}]. The most predominant form of sensory feedback is visual although there are other VR experiences that are based exclusively on haptic (touch) and aural (spatial audio) experiences. With regards to the scope of this article, we would be discussing only visual sensory feedback. Part of having a virtual experience demands the user being \textit{immersed} via VR apparatus into an alternate reality. In general terms, \textit{immersion} refers to a state of mind, a temporary suspension of disbelief which allows a user, to move at free will, from real to virtual and vice versa. Good novelists exploit this fact to pull readers into their story. But none of this immersion is direct and is often presented from a third person point of view. In VR, however, the effect of entering an alternate reality is rather physical than being purely mental. For example, the process of putting on a Head-mounted display physically separates the peripheral vision of a user from the real to the virtual.

Upon entering a VR scene, a simple approach to providing a direct position and orientation feedback would be to track the user's position and orientation in the physical world and transfer the information in real-time, mapped one-to-one inside the VR scene.  There are existing hardware and software tools that enable an exact sync between a user's physical position and orientation to that inside the VR scene by leveraging motion based tracking systems. For the scope of this article, we shall focus on camera based optical tracking systems such as the 5-camera NaturalPoint Optitrack optical tracking system [\autoref{fig:trackingvel}].

\section{Problem Setting}
Suppose a single retro-reflective sphere (similar to those used in motion capture systems) is added to the top of a head-mounted display. Now, suppose we walked into a standard rectangular room with dimensions let us say [16 x 20 x 10]. Inside that room is a camera-based sensing system that reports the 3D position of all reflective infrared spheres in the room relative to center of the floor of the room, that is, it would report [0, 0, height] if the sphere were at the center of the room.  The reports are updated available every 10 millisecond.  We assume that we do not have any way to communicate with the tracking system, but have access to all the data it collects.  There are no guarantees that i) the order of each sphere reported is the same, and ii) that there will be any particular number of spheres in the room at any given time.

Our goal is to integrate the existing tracking system to provide one-to-one direct locomotion control (i.e. real walking) inside a virtual environment for any number of users in the room. We will be considering issues of identification of each user and continuous tracking to achieve seamless transition.

\section{Observations}
\begin{figure}[h]
 \centering
 \includegraphics[width=\columnwidth]{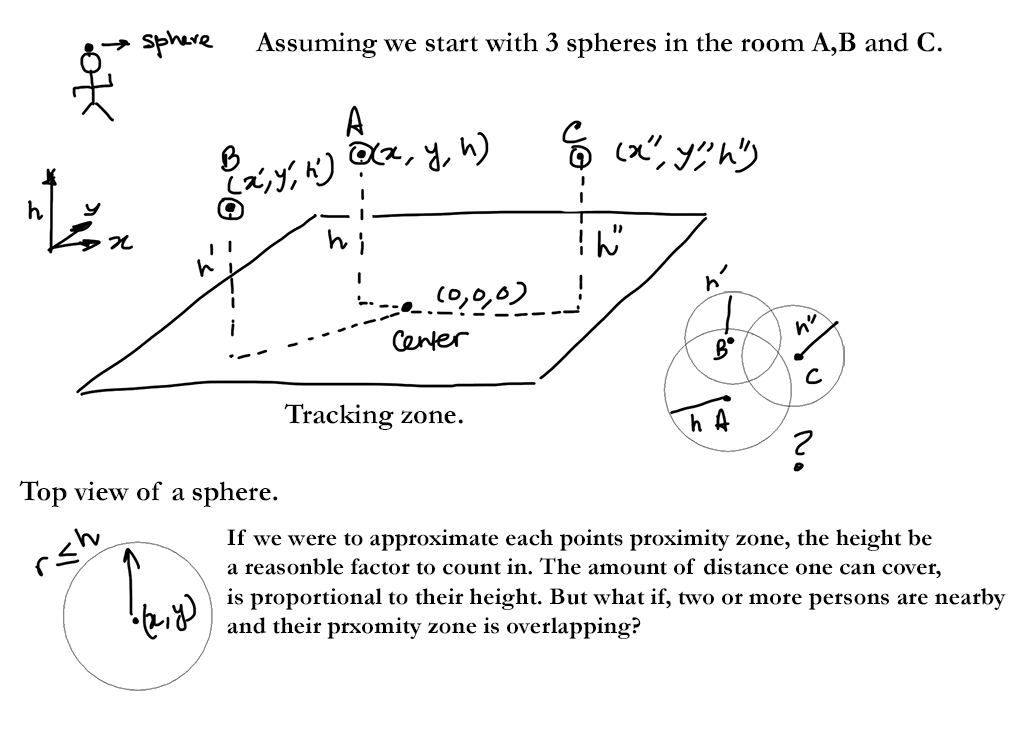}
 \caption{Problem scenario.}
 \label{fig:ProblemScenario}
\end{figure}

We have a single retro-reflective sphere attached to a portable untethered head-mounted display. Let us assume a standardized setup of 5 infrared tracking cameras. From the tracking system’s perspective, we are observing the following:

\begin{enumerate}%[noitemsep]
\item A single retro-reflective sphere is present in the tracking space at the moment to begin with. We start with a single user.
\item We can calculate the height of the present user(s), as the tracking volume is [16 x 20 x 10], by subtracting the height from 10 in the data [0, 0, height], assuming the user is at the center of the room. The height remains constant.
\item The update rate of the data is 10 millisecond.
\item We are assuming that we have access to the tracking data on remote machine that we can access through virtual reality peripheral network (VRPN) \cite{taylor2001vrpn}. 
\end{enumerate}

\section {Solution}

To integrate a robust locomotion scheme for the given setup, we need to solve the following: 1) Figuring out which retro-reflective sphere is which frame to frame, and 2) Figuring out which retro-reflective sphere belongs to every user from their perspective. Above is an illustration of the above problem [\autoref{fig:ProblemScenario}].

In light of the issues illustrated above, we need a calibration step that establishes a direct link between any user(s) and their subsequent movement, which then translates into a one-to-one direct locomotion inside the simulation.

\begin{figure}[h]
 \centering
 \includegraphics[width=\columnwidth]{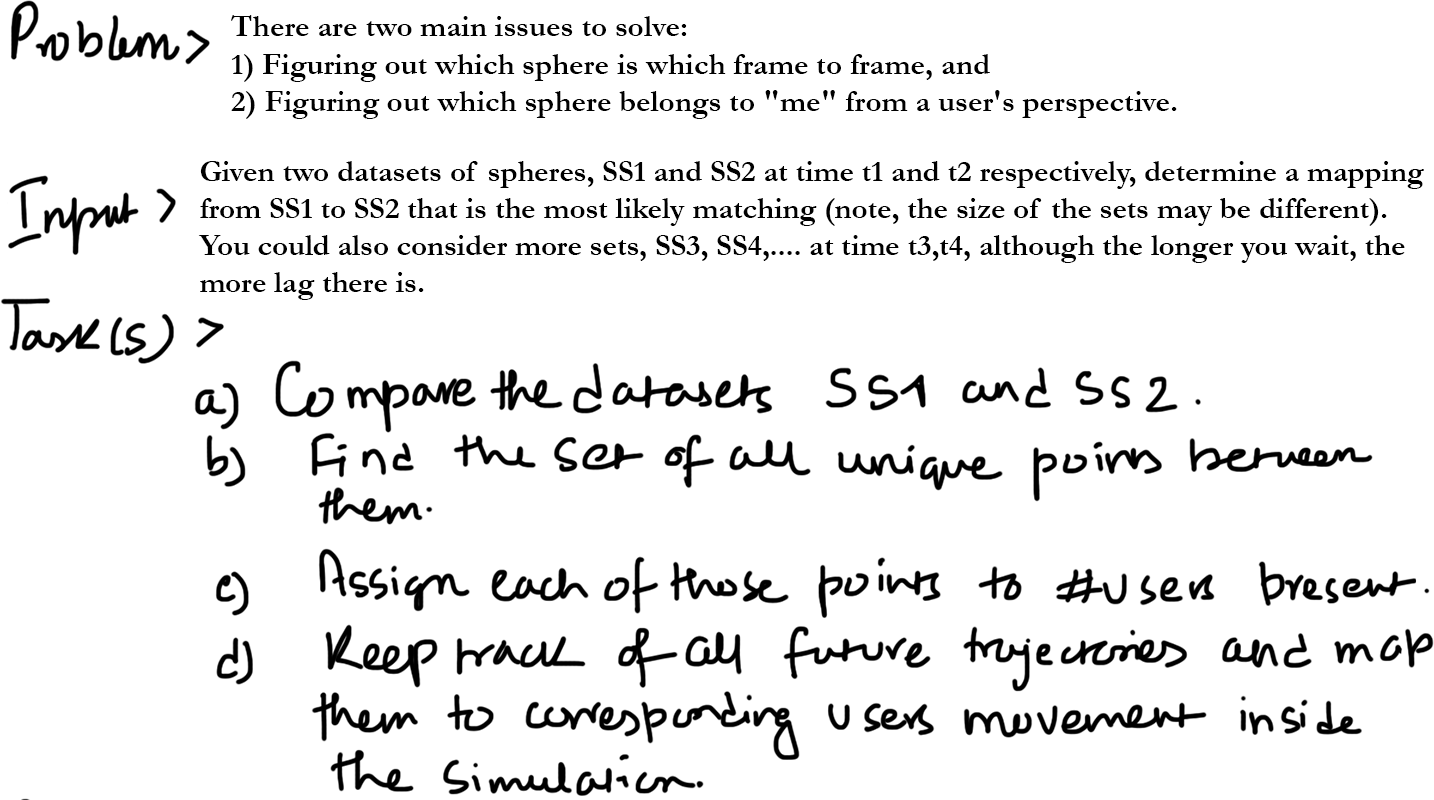}
 \caption{Setting up state prediction and sensor prediction equation.}
 \label{fig:StatePrediction_1}
\end{figure}

\begin{figure}[H]
 \centering
 \includegraphics[width=\columnwidth]{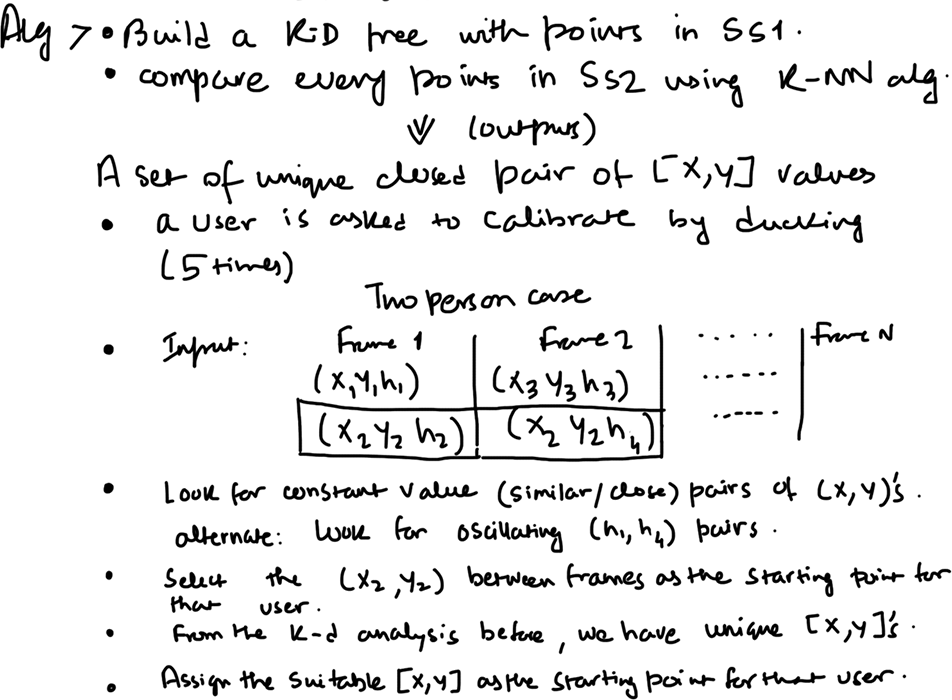}
 \caption{Continued state prediction setup and sensor prediction equation.}
 \label{fig:StatePrediction_2}
\end{figure}

\subsection {Position accuracy assessment}
The assessment of position accuracy of any moving retro spherical point inside the tracking volume needs to be established before applying advanced filtering techniques. Each tracked sphere reports back a vector [x, y, h] every frame. Matching two or temporal sequences of unequal number of motion vectors can do the assessment of position accuracy for each point. The objective here is to find the best alignment among the vector elements using a non-linear alignment method. This task is not trivial, as the number of possible correspondences increases exponentially with the length of compared sequences. A feasible solution to this problem is to implement an algorithm incorporating a calibration step unique for every user. The idea here is to inject a motion behavior that can be used as a signature to establish the mapping between the camera data and the user's trajectory.

\subsection {Algorithm}
We take the first two successive frame-to-frame datasets of tracked points and figure out the unique pairs of points by using k-d trees. This gives you the list of all the starting points for all the users being tracked. The user has been asked to duck for five times before starting to walk. We look for this alternating behavior in the tracked datasets and assign the suitable starting co-ordinates for that user. Finally, we get the entire trajectory for each user through those frames. The following captures [\autoref{fig:StatePrediction_1}, \autoref{fig:StatePrediction_2}] the steps involved in the mapping algorithm.

\begin{figure}[H]
 \centering
 \includegraphics[width=\columnwidth]{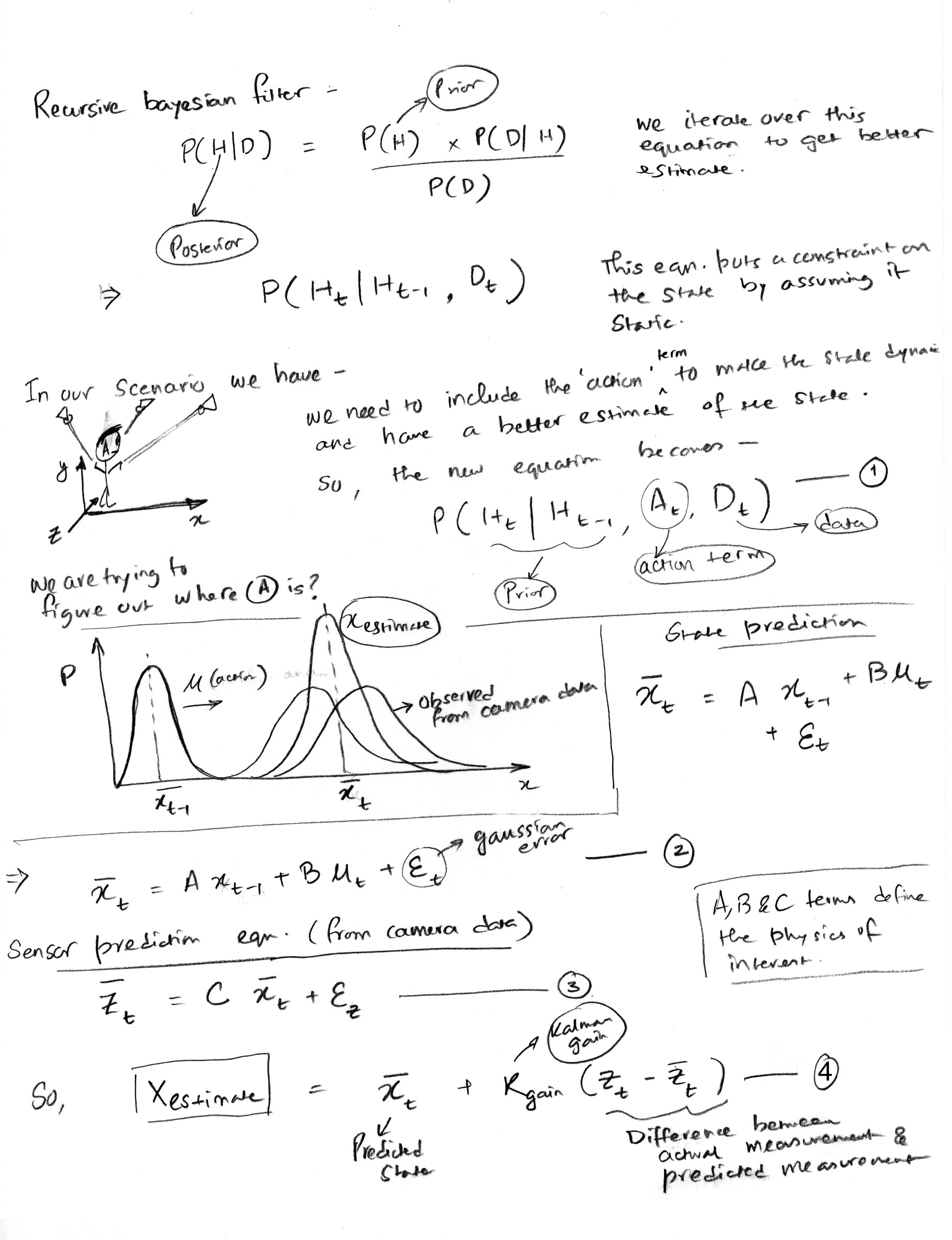}
 \caption{Continued state prediction setup and sensor prediction equation.}
 \label{fig:StatePrediction_3}
\end{figure}

\begin{figure}[H]
 \centering
 \includegraphics[width=\columnwidth]{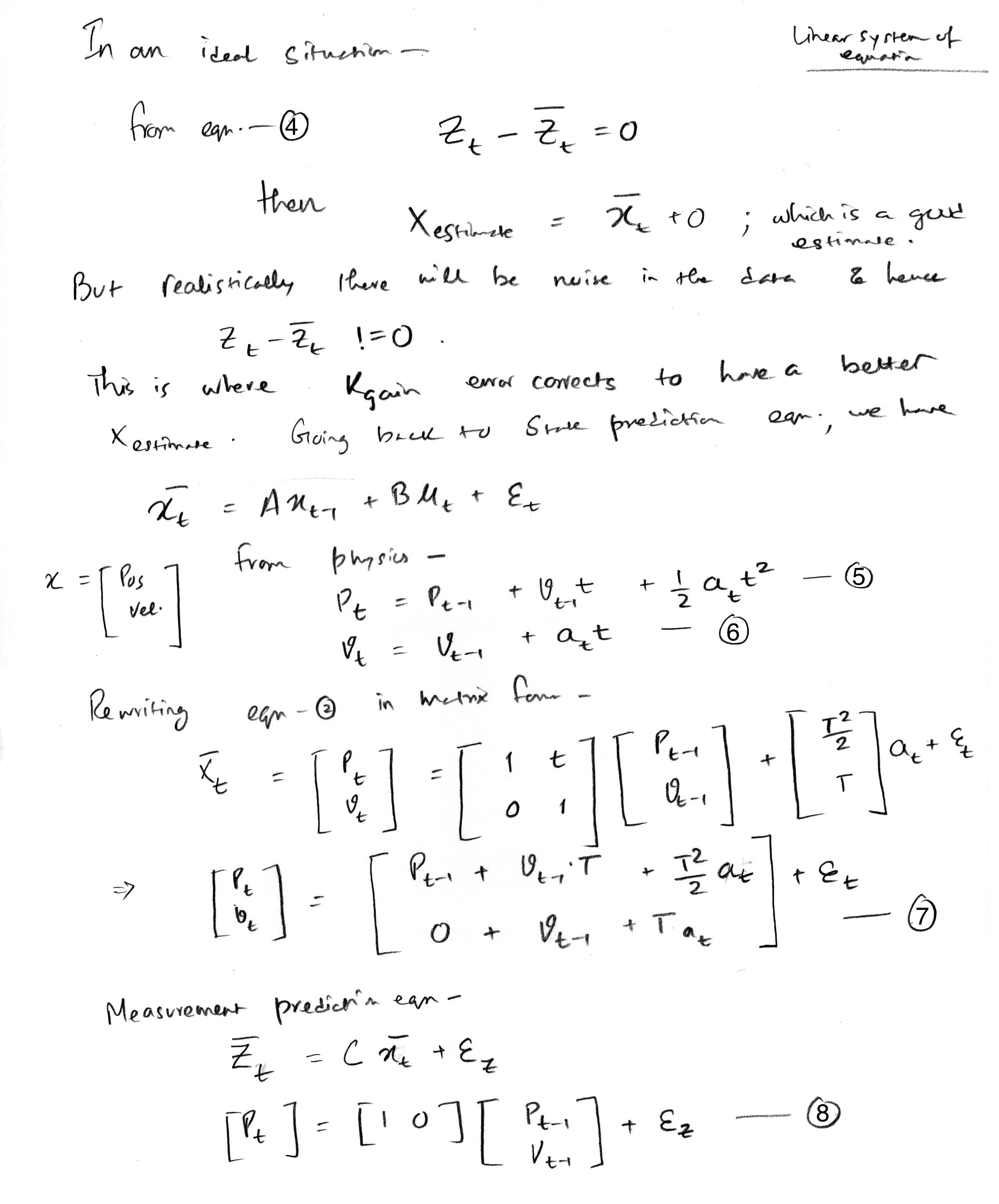}
 \caption{Covariance matrix equation.}
 \label{fig:CovMatrix}
\end{figure}

\begin{figure}[H]
 \centering
 \includegraphics[width=\columnwidth]{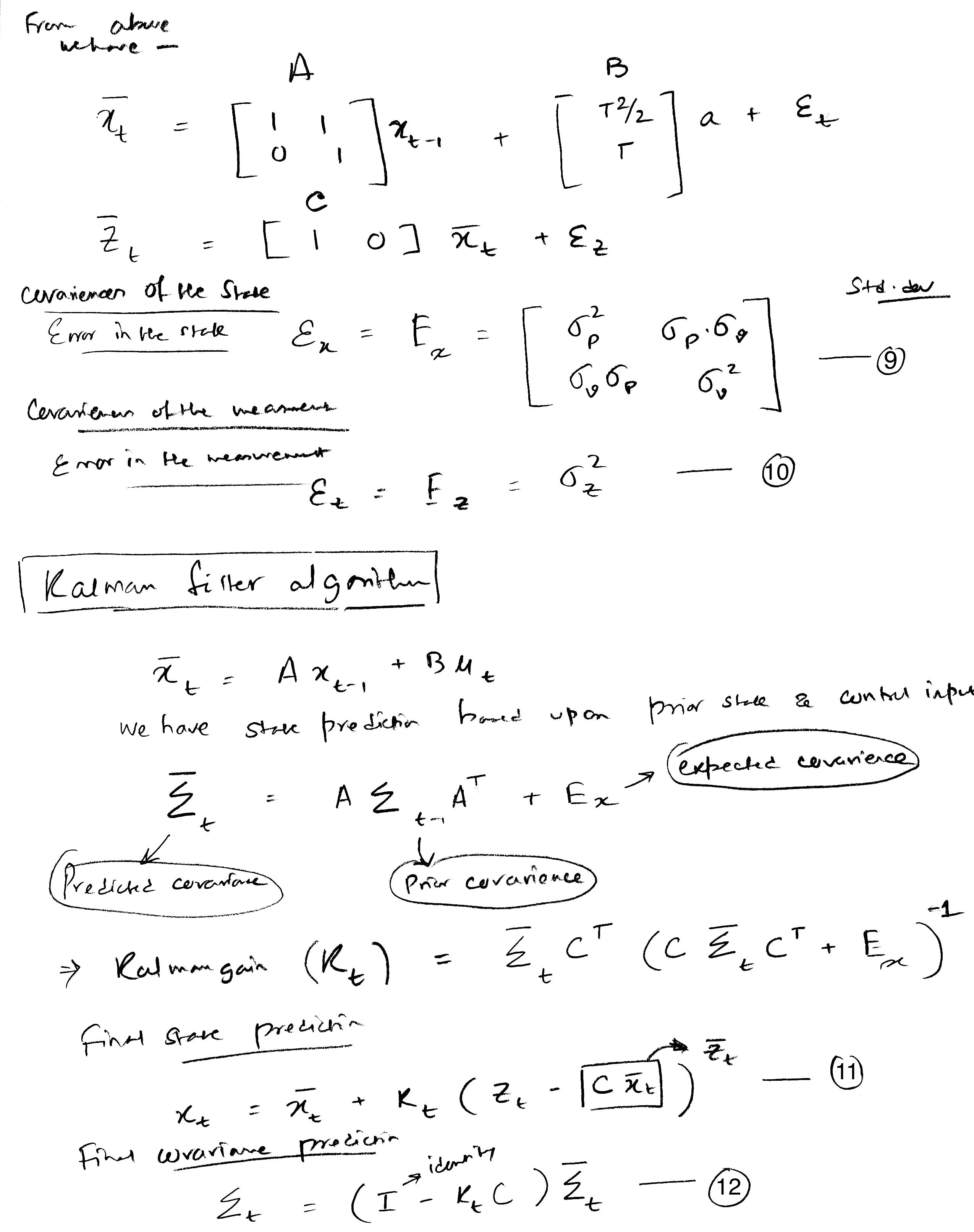}
 \caption{Kalman filter equation.}
 \label{fig:KalmanFilter}
\end{figure}

\subsection{Kalman-gain approach}
The initial approach does not include error correction. To address the issues presented in our previous approach, we look into Kalman filters \cite{kalman1960new} and try to implement that into our system to have better prediction over tracked states for multiple users. From literature, ``The Kalman filter is very powerful in several respects: it supports estimations of past, present, and future states, and it can do so even when the state elements are hidden (not directly observable), or the precise nature of the modeled system is unknown'' \cite{welch2009history}.

The Kalman filter is a specific case of the Bayesian filter equation and we start with a quick look into the Bayesian filter equation and how we can use it recursively to describe our state using the posterior information. The Bayesian filter helps us estimate the posterior given the prior hypothesis and data. But this equation assumes the state of the system to be constant. We need to modify it fit our model of moving targets. The idea here is to map the measurements to a state and have a state update equation. We correct the predicted state based on incoming measurements from the camera data. Equation 2 in [\autoref{fig:StatePrediction_3}] is a modified Bayesian filter that provides a better estimation of the state of the user based on its prior state and the action that was taken to move forward in the virtual world. The error term in equation 2 is assumed to be of type Gaussian and the predicted state is expressed as a linear function of the prior state and the action term to move forward (A, B terms in the equation). These assumptions are there to keep the prediction system optimal. The other part of the overall Kalman filter equation (Equation 4 in [\autoref{fig:StatePrediction_3}]) is the sensor prediction equation (Equation 3 in [\autoref{fig:StatePrediction_3}]). This equation takes the predicted state and transforms it into measurement prediction based on the camera data. This is also a linear system of equation with Gaussian error term in it. The overall idea of Kalman filter is to express the estimated state as a linear function of the predicted state and the difference between the actual measurement and the predicted measurement with the Kalman gain factor acting as the error correcting term.

In our case, the idea is that we map measurements to the state, and we have a state update equation.  We then predict the forward movement of a user in the simulation as needed, and then when a measurement comes in, we correct the prediction. We assume that the state update map and the measurement map are linear and that the noise is white.

\section {Future Work}
This article is purely theoretical and needs experimental validation. The next logical step is to implement the above algorithm in a real-time game engine ecosystem such as Unity\textsuperscript{TM} and validate its efficacy.

\section {Acknowledgment}
This article has stemmed from my Ph.D. comprehensive examination under the supervision of Dr. Kyle Johnsen at the University of Georgia.

%----------------------------------------------------------------------------------------
%	REFERENCE LIST
%----------------------------------------------------------------------------------------

\bibliographystyle{acm}
\bibliography{references}

%----------------------------------------------------------------------------------------

\end{document}